\documentclass[times,twocolumn,final]{elsarticle}

\usepackage{medima}
\usepackage{framed,multirow}

\usepackage{amssymb,amsmath}
\usepackage{xcolor,colortbl}
\usepackage{latexsym}
\usepackage[official]{eurosym}

\PassOptionsToPackage{hyphens}{url}\usepackage{hyperref}\usepackage{url}
\usepackage{xcolor}

\def\highlightnewchanges{0}  

\if\highlightnewchanges1
    \newcommand{\revised}[1]{{\color{blue}#1}}
    \newcommand{\final}[1]{{\color{blue}#1}}
    \newcolumntype{a}{>{\color{blue}}c}
\else
    \newcommand{\revised}[1]{#1}
    \newcommand{\final}[1]{#1}
    \newcolumntype{a}{c}
\fi

\definecolor{newcolor}{rgb}{.8,.349,.1}

\journal{Medical Image Analysis}

\begin{document}

\verso{G. Bortsova, C. Gonz\'{a}lez-Gonzalo, S. C. Wetstein \textit{et~al.}}

\begin{frontmatter}

\title{Adversarial Attack Vulnerability of Medical Image Analysis Systems: Unexplored Factors}

\author[1]{Gerda \snm{Bortsova}\corref{cor1}\fnref{fn1}}
\cortext[cor1]{Corresponding author.}
  \ead{gerdabortsova@gmail.com}
\author[2,3]{Cristina \snm{Gonz\'{a}lez-Gonzalo}\corref{cor1}\fnref{fn1}}
\fntext[fn1]{The first three authors contributed equally to this work.}
\author[4]{Suzanne C. \snm{Wetstein}\corref{cor1}\fnref{fn1}}
\author[1]{Florian \snm{Dubost}}

\author[5]{Ioannis \snm{Katramados}}

\author[5]{Laurens \snm{Hogeweg}}

\author[2,3]{Bart \snm{Liefers}}

\author[6]{Bram \snm{van Ginneken}}

\author[4]{Josien P.W. \snm{Pluim}}

\author[4]{Mitko \snm{Veta} \fnref{fn2}}
\fntext[fn2]{The last three authors contributed equally to this work.}
\author[2,3,7]{Clara I. \snm{S\'{a}nchez}\fnref{fn2}}

\author[1,8]{Marleen \snm{de Bruijne}\fnref{fn2}}

\address[1]{Biomedical Imaging Group Rotterdam, Department of Radiology and Nuclear Medicine, Erasmus MC, The Netherlands}
\address[2]{A-Eye Research Group, Diagnostic Image Analysis Group, Department of Radiology and Nuclear Medicine, Radboudumc, Nijmegen, The Netherlands}
\address[3]{Donders Institute for Brain, Cognition and Behaviour, Radboudumc, Nijmegen, The Netherlands}
\address[4]{Medical Image Analysis Group, Department of Biomedical Engineering, Eindhoven University of Technology, Eindhoven, The Netherlands}
\address[5]{Intel Corporation, The Netherlands}
\address[6]{Diagnostic Image Analysis Group, Department of Radiology and Nuclear Medicine, Radboudumc, Nijmegen, The Netherlands}
\address[7]{Department of Ophthalmology. Radboudumc, Nijmegen, The Netherlands}
\address[8]{Department of Computer Science, University of Copenhagen, Denmark}

\begin{abstract}
Adversarial attacks are considered a potentially serious security threat for machine learning systems. Medical image analysis (MedIA) systems have recently been argued to be vulnerable to adversarial attacks due to strong financial incentives and the associated technological infrastructure.

In this paper, we study previously unexplored factors affecting adversarial attack vulnerability of deep learning MedIA systems in three medical domains: ophthalmology, radiology, and pathology. We focus on adversarial black-box settings, in which the attacker does not have full access to the target model and usually uses another model, commonly referred to as surrogate model, to craft adversarial examples that are then transferred to the target model. We consider this to be the most realistic scenario for MedIA systems.
Firstly, we study the effect of weight initialization (pre-training on ImageNet or random initialization) on the transferability of adversarial attacks from the surrogate model to the target model, i.e., how effective attacks crafted using the surrogate model are on the target model. Secondly, we study the influence of differences in development (training and validation) data between target and surrogate models. 
We further study the interaction of weight initialization and data differences with differences in model architecture. All experiments were done with a perturbation degree tuned to ensure maximal transferability at minimal visual perceptibility of the attacks. 

Our experiments show that pre-training may dramatically increase the transferability of adversarial examples, even when the target and surrogate's architectures are different: the larger the performance gain using pre-training, the larger the transferability.
Differences in the development data between target and surrogate models considerably decrease the performance of the attack; this decrease is further amplified by difference in the model architecture.
We believe these factors should be considered when developing security-critical MedIA systems planned to be deployed in clinical practice. We recommend avoiding using only standard components, such as pre-trained architectures and publicly available datasets, as well as disclosure of design specifications, \revised{in addition to using adversarial defense methods}. When evaluating the vulnerability of MedIA systems to adversarial attacks, various attack scenarios and target-surrogate differences should be simulated to achieve realistic robustness estimates.

\final{The code\textsuperscript{3} and all trained models\textsuperscript{4} used in our experiments are publicly available.}
\fntext[fn3]{\final{\url{https://github.com/Gerda92/adversarial_transfer_factors}}}
\fntext[fn4]{\final{\url{https://doi.org/10.5281/zenodo.4792375}}}

\end{abstract}


\end{frontmatter}


\section{Introduction}
Deep learning (DL) has been shown to achieve close or even superior performance to that of experts in many medical image analysis (MedIA) applications, including in ophthalmology \citep{gulshan2016development, ting2017development}, radiology \citep{rajpurkar2017chexnet}, and pathology \citep{bejnordi2017diagnostic, bulten2020automated, wetstein2020deep}. This has created an opportunity to automate certain medical tasks and integrate DL systems in clinical settings \citep{abramoff2018pivotal, murphy2020computer, GE_FDA}. \revised{However,} a threat to DL systems is posed by so-called \emph{adversarial attacks} \citep{szegedy2013intriguing}. Such attacks apply a carefully engineered, subtle perturbation to the input of the target model to cause misclassification. Those perturbed inputs, referred to as \emph{adversarial examples}, have been shown effective in forcing state-of-the-art systems to produce incorrect predictions \citep{goodfellow2014explaining, madry2017towards}. 

\revised{Adversarial attacks are not the only kind of malicious manipulation of input to DL models that changes their predictions. Adversarial attacks are manipulations that aim to preserve the semantic contents of a given image, e.g., whether it is healthy or diseased, while changing the prediction of the network for the image. Apart from this type of attack, images can also be manipulated to change their content: for example, signs of disease can be removed from a diseased image or added to a healthy image \citep{xia2020pseudo, sun2020adversarial, baumgartner2018visual, becker2019injecting}, which, in turn, can change network predictions. However, developing these synthetically changed images remains challenging \citep{xia2020pseudo}, as it is hard to guarantee they look realistic, hard to control which image structures are changed, and these algorithms may be difficult to train and require large training datasets. In contrast, adversarial examples generated by adding noise of bounded, small magnitude are guaranteed to look realistic and do not induce any unpredictable changes in the image. Therefore, we consider adversarial attacks to be a more feasible, and thus more likely type of attack on MedIA systems, which is why we have limited our scope to adversarial attacks.}

\subsection{\revised{Context of adversarial attacks in MedIA}}
\revised{A recent market report has predicted that through 2022, 30\% of all cyberattacks against systems powered by artificial intelligence (AI) will leverage training-data poisoning, AI model theft, or adversarial examples \citep{cearly2019gartner}. This results especially alarming for the healthcare industry, considering that it is predicted to suffer two to three times more cyberattacks than the average amount for other industries \citep{cisco2019}. Limited resources and fragmented governance on cybersecurity \citep{martin2017cybersecurity, ghafur2019challenges}, and larger consequences at both financial \citep{IBM2020} and human levels \citep{martin2017cybersecurity} make healthcare particularly vulnerable to cyberattacks.}

Adversarial attacks may \revised{therefore} pose a large threat in the medical domain \citep{finlayson2019adversarial, finlayson2018adversarial}. This is due to two main factors: financial interests and technical sources of vulnerability. 

Firstly, some parties involved in healthcare systems have a financial interest in manipulating patient diagnosis and prognosis. \revised{Healthcare fraud has been shown to be committed by large companies as well as individuals \citep{rudman2009healthcare, kalb1999health}. When expressed as a proportion of the global healthcare expenditure estimated by the World Health Organisation in 2013 (\$7.35 trillion or \euro5.65 trillion), the global average healthcare fraud and error loss equates to 6.19\% (\$455 billion or \euro350 billion) \citep{gee2015financial}.} In the future, adversarial attacks could be used as a tool to manipulate MedIA systems supporting insurance, clinical, or drug/device approval decisions. Adversarial attacks can boost existing fraudulent behavior in fee-for-service healthcare systems, such as the one in the United States, where healthcare providers and insurance companies manipulate diagnostic codes of patients to affect reimbursement decisions. Fraudulent behavior involving adversarial attacks could potentially be more difficult to detect compared to manipulating diagnostic codes directly. Adversarial attacks can also be used to bias patient diagnosis towards false referrals or unnecessary prescriptions of medication or treatment. Similarly, companies could bias trial outcomes and gain the favor of regulatory bodies, such as the United States Food and Drug Administration, by showing the desired effect of a drug/device to be approved. \revised{It is important to emphasize that these attacks would be facilitated because the attacker would be already inside the healthcare infrastructure.} These situations can result in deteriorated quality of healthcare, financial loss, decreased trust in MedIA systems and hence impediments to their integration into clinical practice.

The second factor that may facilitate adversarial attacks in the medical domain concerns technical sources of vulnerability. These include domain-specific characteristics of medical images, such as highly-standardized image acquisition protocols, and the security of technological infrastructure into which MedIA systems will be embedded \citep{ma2021understanding, finlayson2019adversarial}. \revised{In this case, the attacks would be performed most commonly from outside the healthcare infrastructure, by means of a breach. In a recent investigation, more than 45 million medical images and their patient metadata were found to be exposed and freely accessible, without hacking tools required, on over 2,000 unprotected medical servers across 67 countries, including the United States, United Kingdom, France, and Germany \citep{cybelangel2020}. A survey from 2017 revealed that healthcare data breaches have affected one in four consumers in the United States \citep{accenture2017}. The security risks of such breaches include blackmail and ransomware \citep{forbes2021}, as well as malicious data manipulation. Among deployed connected medical devices, imaging systems (including systems for image acquisition, viewers, workstations, and servers) have been found to have the most security issues, mainly derived from user practice and outdated infrastructure \citep{hcinnovation2018}. This last aspect is strongly related to widely used software and protocols, such as DICOM, which were developed before cybersecurity was a concern and leave serious security gaps \citep{eichelberg2020cybersecurity, stites2016secure}.}

\subsection{\revised{Adversarial attacks and defenses}}
Multiple methods to generate adversarial attacks have been proposed in the literature and can be categorized following different taxonomies \citep{yuan2019adversarial, akhtar2018threat,biggio2018wild}. As an example, some methods perform one-shot attacks \citep{szegedy2013intriguing, goodfellow2014explaining}, whereas other methods optimize the attack in an iterative way \citep{madry2017towards, kurakin2016adversarial, papernot2016limitations, moosavi2016deepfool, carlini2017towards, su2019one, moosavi2017universal}. Similarly, there are methods that generate a specific perturbation for each input \citep{szegedy2013intriguing, goodfellow2014explaining, madry2017towards, kurakin2016adversarial, papernot2016limitations, moosavi2016deepfool, carlini2017towards, su2019one} and methods that generate universal perturbations that can be applied to any image \citep{moosavi2017universal, brown2017adversarial}.

Furthermore, adversarial attack methods can be applied in scenarios with different degrees of knowledge of the target system: from having full knowledge (\emph{white-box attacks}) \citep{goodfellow2014explaining} to being agnostic to the (hyper)parameters of the target model (\emph{black-box attacks}) \citep{papernot2017practical}. The latter usually use another model, commonly referred to as \emph{surrogate model}, to craft adversarial examples that are then transferred to the target model. The effectiveness of a black-box attack is determined by its \emph{transferability} between the surrogate model and the target model \citep{papernot2017practical}.

Several studies have investigated the impact of adversarial attacks on MedIA systems specifically. This has been studied for classification and segmentation problems in different imaging modalities, including color fundus imaging \citep{finlayson2018adversarial, ma2021understanding, ozbulak2019impact}, chest X-ray \citep{finlayson2018adversarial, taghanaki2018vulnerability, ma2021understanding}, dermoscopy \citep{finlayson2018adversarial, ma2021understanding, paschali2018generalizability, ozbulak2019impact}, and brain MRI \citep{paschali2018generalizability}. In these studies, adversarial attacks were proven effective in both white- and black-box settings.

Correspondingly, numerous defense methods  \citep{yuan2019adversarial, papernot2017practical,biggio2018wild} have been proposed to protect DL systems from adversarial attacks by training networks so as to robustify them against adversarial attacks \citep{goodfellow2014explaining, papernot2016distillation} or by detecting adversarial examples or neutralizing adversarial noise \citep{lu2017safetynet, song2017pixeldefend}. Defense methods have also been considered to protect MedIA systems from adversarial attacks \citep{ma2021understanding}.
Nevertheless, almost all proposed countermeasures have been shown to be only effective against some attacks \citep{yuan2019adversarial}, hardly work against infinitesimal perturbations \citep{papernot2017practical}, or can easily be made ineffective if the attacker is aware of them
\revised{\citep{uesato2018adversarial, athalye2018obfuscated, carlini2017adversarial}}.

\subsection{\revised{Adversarial vulnerability of MedIA systems}}
A better understanding of factors affecting the vulnerability of MedIA systems is therefore crucial to inform and improve the evaluation of their robustness against adversarial attacks, as well as the design of new MedIA systems.
There are several factors related to the design of the target model, such as network architecture \citep{szegedy2013intriguing, su2018robustness}, and the attack scenario, such as disparity in the development data, i.e., difference in the data used for training and validation, between the target and the attacker \citep{szegedy2013intriguing}, that affect the transferability of adversarial attacks and thus the vulnerability of the systems. Although factors such as network architecture disparity (i.e. having different network architectures) \citep{paschali2018generalizability, taghanaki2018vulnerability} are sometimes considered when evaluating vulnerability of MedIA systems against adversarial attacks, the impact of other crucial aspects of real-world MedIA scenarios has not been explored yet.

In this paper, we focus on two unexplored factors that can potentially influence adversarial attack transferability in MedIA systems: ImageNet pre-training and development data disparity.
The key contributions of our paper are:
\begin{itemize}
\item We study the effect of ImageNet pre-training on adversarial attack transferability. Since systems pre-trained on natural images have shown to achieve improved performance in shorter training times in several medical applications \citep{gulshan2016development, wang2017chestx}, pre-training on ImageNet has become a common design choice for development of MedIA systems \citep{litjens2017survey}. Pre-trained models may be more similar to each other compared to randomly initialized models due to retaining information learned from ImageNet. However, to the best of our knowledge, no studies (of MedIA or any other DL systems) have compared transferability of adversarial attacks between pre-trained models to that between randomly initialized models.
\item We study the effect of disparity in the data used for development of the target and surrogate models. With increasing availability of high-quality, large public datasets, it becomes more likely that MedIA systems will, at least partly, rely on these easily accessible data in order to fulfill the requirement of large datasets for DL development. Simultaneously, MedIA systems in the deployment stage might also make use of larger amounts of private data \citep{abramoff2016improved, gonzalez2020evaluation, murphy2020computer}. Comparing adversarial transferability in scenarios of development data parity and disparity may provide further insight on how vulnerable MedIA systems are. \revised{Additionally, we study adversarial robustness of ImageNet pre-trained and randomly initialized networks trained using smaller development sets under an attack scenario of data disparity, simulating target models developed with small, private datasets.
}

\item We investigate these factors in three popular medical applications: detection of referable diabetic retinopathy in color fundus images, classification of pathologies in chest X-Ray, and breast cancer metastasis detection in histological lymph node sections.
\end{itemize}

We used the following methodology to study the effect of ImageNet pre-training and development data disparity on adversarial transferability.
We implemented different adversarial attack methods and applied them to different state-of-the-art network architectures, which allows us to additionally evaluate the effect of network architecture disparity: \revised{i.e., the effect of target and surrogate models having a different architecture as compared to them having the same architecture}. We perform our experiments in varying black-box settings, which we consider to be the most realistic attack scenario for MedIA systems.
In contrast to previous studies, we analyze and adjust the perturbation degree used in our experiments so as to ensure optimal transferability at minimal visual perceptibility of the adversarial attacks, considering human input is often required in MedIA settings. 
We thoroughly examine the implications of our results on the design of MedIA systems, as well as provide recommendations for evaluating their robustness against adversarial attacks.

\section{Related work}

Black-box attacks can have varying degrees of interaction with the target model: from having no interaction at all to unlimited ‘querying’  of the model and using its predictions in crafting adversarial perturbations (for example, one-pixel attacks by \cite{su2019one}, or oracle attacks such as the one proposed by \cite{papernot2016transferability}).
The non-query-based type of black-box attacks is the focus of this work and is, perhaps, the most commonly studied \citep{akhtar2018threat,yuan2019adversarial}, including in the MedIA field \citep{finlayson2018adversarial,paschali2018generalizability,taghanaki2018vulnerability}.

Black-box attacks that do not allow querying the target model typically rely on the transferability of adversarial perturbations from a surrogate model to the target model. Adversarial examples have been shown to be transferable between highly distinct models \citep{szegedy2013intriguing, liu2016delving, moosavi2017universal}.
The transferability of adversarial examples between different models can be explained by the similarity of their decision boundaries \citep{tramer2017space} and depends on how similar their design and training are \citep{uesato2018adversarial}. 
Perhaps, the most well-studied factor affecting adversarial transferability is disparity in model architecture \citep{su2018robustness}. Relatively few studies have investigated the influence of other kinds of target-surrogate differences on the success of adversarial attacks: most studies trained their target and surrogate models on exactly the same subset of the same dataset, and use the same pre-processing, data augmentation, weight initialization, training loss function, and other training parameters.

In this study, we focus on the effects of two previously unexplored factors in MedIA settings on the transferability of black-box attacks: pre-training on ImageNet and disparity in the development data between target and surrogate models. We also study the interaction of both factors with network architecture disparity. Below we provide an overview of the literature related to these factors:

\textbf{Pre-training on ImageNet.}
In the MedIA field, DL methods commonly use pre-training on natural images to improve performance  \citep{litjens2017survey}.
Pre-trained networks have also often been used in studies on adversarial robustness \citep{finlayson2018adversarial,paschali2018generalizability,ma2021understanding}.
However, all studies either considered target and surrogate models that were both pre-trained or both randomly initialized. To our knowledge, no studies have compared adversarial attack transferability between DL networks pre-trained on ImageNet (or any other dataset used for performance boosting) to that between randomly initialized networks. We hypothesize that the transferability of adversarial examples between pre-trained target and surrogate models may be larger than that between randomly initialized models, since pre-training might increase the similarity between models due to retaining information learned from ImageNet.

\revised{Although the effect of ImageNet pre-training was not studied in the black-box attack scenario, its effect on white-box adversarial robustness was studied by \cite{hendrycks2019using} for networks that were adversarially fine-tuned: i.e., trained using adversarial training \citep{madry2017towards} on the target data after pre-training. In their study, regular ImageNet pre-training had no positive effect on the white-box robustness of networks adversarially fine-tuned on CIFAR. However, adversarial ImageNet pre-training increased the robustness substantially. The effect on robustness of adversarial pre-training for networks that are fine-tuned normally (not adversarially) was not reported by \cite{hendrycks2019using} or others.}

\textbf{Disparity in development data.}
\cite{szegedy2013intriguing} reported that adversarial examples crafted using a surrogate model trained on a different (similarly sized) data subset as the target model are substantially less transferable than those crafted using the same training data for the target and surrogate model.
However, they only demonstrated this for simple fully-connected models trained on MNIST.
No further studies have focused on the effect of training data disparity, including in the MedIA field: all studies of black-box attacks on MedIA DL assumed perfect data parity \citep{finlayson2018adversarial,paschali2018generalizability,taghanaki2018vulnerability}.
This factor is particularly important to study in the context of MedIA systems, where some systems are trained on easily accessible public data, whereas others rely on private data.
In the case of using only public data for development, we can assume that surrogate models can be trained with the same dataset as the target (data parity), while in the case of using private data, this is not possible (data disparity). We believe it is important to consider these different scenarios and study the influence of data (dis)parity on transferability of adversarial examples in MedIA systems.

\textbf{Disparity in model architecture.}
\cite{su2018robustness} studied the adversarial robustness of 18 well-known image classification models trained on ImageNet. Their findings suggest that adversarial examples crafted from one model can only be transferred within the same family (e.g. VGGs or Densenets). They also found that deeper models within the same family are slightly more robust than shallower models, but differences in model architecture were found to affect transferability more than differences in model size.
There have been no similarly comprehensive studies on architecture disparity or adversarial example transferability between different architectures for MedIA systems.
However, some studies reported attack performance under both architecture parity and disparity \citep{paschali2018generalizability} or under disparity only \citep{taghanaki2018vulnerability}. \cite{szegedy2013intriguing} found that having architecture disparity in addition to development data disparity further reduced the transferability of attacks. In this study, we investigate the interaction of architecture (dis)parity with weight initialization (pre-training on ImageNet or random initialization) and development data (dis)parity.

\section{Methods}
\subsection{Threat model}
The security of any system is measured in relation to the capabilities and goals of its potential adversaries. The limits to the attacker’s capabilities, including their knowledge, and their goals are captured by the concept of a \emph{threat model}. In the context of evaluating adversarial robustness of machine learning systems, explicitly specifying the considered threat model helps to clearly delineate the scope of attacks against which robustness is studied and thus allows for falsifiable claims \citep{carlini2019evaluating}. The threat model considered in this study is the following:

\textbf{Goal.} We assume the attacker’s goal is to cause general misclassification, which is usually called an \emph{untargeted} adversarial attack. In an untargeted adversarial attack the goal is to modify the input in a way that it will be classified as any class but the ground-truth class, whereas in a targeted adversarial attack the goal is to modify the input in a way that it will be classified as a specific class.

\textbf{Capability.} We assume the attacker's capabilities are:
\begin{itemize}
\item The attacker can only manipulate the input to the target system (we assume this input is directly fed into DL networks) and only at inference time.
\item The attacker is allowed to modify the input images in a way that appears very subtle or even imperceptible to the human eye.
\item The attacker cannot query the target model.
\end{itemize}

\textbf{Knowledge.}
We simulate scenarios of the attacker lacking knowledge of the following features of the target model: weight initialization (pre-trained on ImageNet or randomly initialized), data used for development, and network architecture. The weights of the target model cannot be accessed by the attacker in all attack scenarios we consider.

\subsection{Adversarial attacks}
In this study, we used two adversarial attack methods that were most commonly and effectively used in the literature:
fast gradient sign method (FGSM) \citep{goodfellow2014explaining} and projected gradient descent (PGD) \citep{madry2017towards}.

\textbf{Fast gradient sign method.} FGSM is a one-shot attack method in which the adversarial perturbation is computed as the sign of the gradient of the loss with respect to the input image.
The sign of the gradient in every pixel determines whether $\epsilon$, the parameter regulating the maximum amount of perturbation, is added or subtracted from every pixel in the target image $x$ to create an adversarial example: 
\begin{equation}
x_{adv} = x + \epsilon\cdot sign\big(\nabla_x\mathcal{L}(f(x; \theta), y) \big),
\end{equation}
where $\mathcal{L}$ represents the loss, $f$ the selected network architecture, $\theta$ the corresponding parameters, and $y$ the image label.

\textbf{Projected gradient descent.} PGD is an iterative version of FGSM, in which several steps for computing the perturbation and adding it to the input are performed: 
\begin{equation}
x^{(i+1)}_{adv} = clip_x^\epsilon \big\{ x^{(i)} + \alpha\cdot sign \big( \nabla_x\mathcal{L}(f(x^{(i)}; \theta), y) \big) \big\},
\end{equation}
where $\alpha$ controls the step size; $\epsilon$ is the parameter regulating the maximum degree of perturbation added to every pixel; $clip_x^\epsilon$ function clips its input so that it does not deviate from $x$ more than $\epsilon$ as measured by $\ell_\infty$ norm.

In the black-box setting, $f'(\cdot, \theta')$, where $f'$ is the surrogate network architecture and $\theta'$ are the corresponding parameters, is used to compute the attack, which is then transferred to the target model.

\subsection{Network architectures, training, and data}

We selected Inception-v3 \citep{szegedy2016rethinking} and Densenet-121 \citep{huang2017densely} as the base architectures for our experiments. Both architectures were previously applied in the selected medical applications and achieved good performance \citep{gulshan2016development, rajpurkar2017chexnet, guendel2018learning, veeling2018rotation}.
All networks were trained until convergence on a validation set using Adam optimization with learning rate decay and binary cross-entropy loss.

For the dataset used in each application, a development and a test set were defined. The development set was used for training and validation. The independent test set was used to measure the performance of each model on clean and adversarial examples. We randomly divided all development sets, at patient-level, into two non-overlapping, equal-sized parts --- \emph{d1} and  \emph{d2} --- to be able to study the influence of data parity on attack transferability. 
\revised{Two more sets}, \emph{d2/2} \revised{and \emph{d1/10}, were} created by randomly sampling at patient level half of \emph{d2} \revised{and 10\% of \emph{d1}, respectively. This was done} to study the influence of dataset size. The description of each dataset and dataset-specific network parameters is stated below. Table \ref{datapartitions} provides an overview of data partitioning for each dataset.

\textbf{Ophthalmology.}
We used the Kaggle dataset for diabetic retinopathy detection \citep{kaggle2015diabetic}, which contains 88,702 color fundus images with manually-labeled diabetic retinopathy severity.
In order to have more images available for development, as proposed in Finlayson et al. \citep{finlayson2018adversarial}, we merged the original training (35,126 images) and test sets (53,576 images) and split the images randomly at patient-level subsets for development (88\%) and testing (12\%). 

Pre-processing included extracting the field of view and rescaling to $512\times{512}$ pixels. The networks were trained 
to distinguish between non-referable (stages 0 to 1) and referable diabetic retinopathy (stages 2 to 4) using batch class balancing. For data augmentation, we used flipping and rotation.

\textbf{Radiology.}
We used the ChestX-Ray14 dataset \citep{wang2017chestx}, consisting of 112,120 frontal-view X-rays annotated with 14 non-mutually-exclusive pathology labels. The official data split (80\%-20\%) was used to define our development and test sets.

Pre-processing included downsampling images to $256 \times 256$ resolution.
The architectures were trained using binary cross-entropy loss to predict 14 pathology classes and one ``no finding'' class.
For data augmentation, we used translation and horizontal flipping.

\textbf{Pathology.}
We used the PatchCamelyon (PCam) \citep{veeling2018rotation} dataset, which contains 327,680 patches extracted from histopathology whole-slide images of lymph node sections.
The official data split (90\%-10\%) was used to define our development and test sets.

The networks were trained to distinguish between the presence or absence of metastatic tissue in the patch center. For data augmentation, we used horizontal and vertical flipping and random color augmentations. 

\begin{table*}[h!]
\caption{Size of development and testing subsets for each dataset. In order to study the effect of development data disparity on adversarial attack transferability, development sets were divided into two equal-sized parts: \emph{d1} and \emph{d2}; \revised{\emph{d1} was subsequently subsampled to obtain small datasets which contained 10\% of the data \emph{d1/10};} \emph{d2} was subsequently subsampled to obtain a half-sized subset \emph{d2/2}.
}
\label{datapartitions}
\centering
\begin{tabular}{c|c|c|c|c|c|c|c}
\multicolumn{2}{c|}{}               & \multicolumn{2}{c|}{Ophthalmology} & \multicolumn{2}{c|}{Radiology} & \multicolumn{2}{c}{Pathology}  \\ 
\hline
\multirow{4}{*}{Development} & d1   & \multirow{4}{*}{79,058 (88\%)} & 39,030      & \multirow{4}{*}{86,524 (80\%)} & 43,657  & \multirow{4}{*}{294,912 (90\%)} & 147,456  \\ 
\cline{2-2}\cline{4-4}\cline{6-6}\cline{8-8}
                             & d2   &                           & 39,028      &                           & 42,867  &                           & 147,456  \\ 
\cline{2-2}\cline{4-4}\cline{6-6}\cline{8-8}
                             & \revised{d1/10} &                           & \revised{3,906}      &                           & \revised{3,989} &                           & \revised{14,892}  \\
\cline{2-2}\cline{4-4}\cline{6-6}\cline{8-8}
                             & d2/2 &                           & 19,514      &                           & 23,666 &                           & 73,728  \\ 

\hline
\multicolumn{2}{c|}{Test}           & \multicolumn{2}{c|}{10,644 (12\%)}      & \multicolumn{2}{c|}{25,596 (20\%)}  & \multicolumn{2}{c}{32,768 (10\%)}   \\ 
\hline
\multicolumn{2}{c|}{Total}          & \multicolumn{2}{c|}{88,702 (100\%)}     & \multicolumn{2}{c|}{112,120 (100\%)} & \multicolumn{2}{c}{327,680 (100\%)} 
\end{tabular}
\end{table*}

\section{Experimental setup}
In all experimental setups, the performance of the target models on the test set of each dataset was measured using the area under the receiver operating characteristic curve (AUC) or mean AUC for the multi-class case.

\subsection{Perturbation degree}

Firstly, we analyzed the effect of perturbation degree on the adversarial attacks to ensure maximal transferability at minimal visual perceptibility in further experiments. To our knowledge, only one study has systematically analyzed the effect of perturbation degree in MedIA settings \citep{ma2021understanding}, although it was only done for white-box attacks. We believe perturbation degree is a parameter that should be further investigated to yield more accurate estimations of robustness against adversarial attacks. In this study, we analyzed the performance of FGSM and PGD attacks and the visual perceptibility under different degrees of perturbation, controlled by $\epsilon$: \revised{0.01, 0.02, 0.03, 0.04, 0.05, and 0.06}. These values were applied to image intensities rescaled between -1 and 1.
\revised{We assessed visual perceptibility of attacks bounded by different epsilons in two different ways. Firstly, the first authors used their own visual perception to judge how subtle adversarial perturbations appear when adversarial and original images were viewed in juxtaposition. Due to impracticality of assessing every adversarial input to our models, this was evaluated in a subset of images of each modality and for each epsilon. Secondly, we computed mean Structural Similarity Index Measure (SSIM) \citep{wang2004image} between adversarial and original versions of all images for each modality and epsilon. SSIM is based on a hypothesized characteristic of the human visual system to be sensitive to structural information in images and was previously shown to be a robust measure of perceptual quality of images \citep{wang2004image}.}

For the PGD attacks, we used step size
$\alpha = 0.01$ and 20 iterations. In this experiment, all models were randomly initialized and trained on the same partition of the development set, \emph{d1}.

To ensure that the decrease in target model performance after an adversarial attack is due to the adversarial nature of the perturbation and not solely due to added noise, we additionally computed ``control'' noise. While existing works chose standard noise distributions such as Gaussian for this purpose \citep{paschali2018generalizability}, we chose to compare adversarial perturbations with their randomly spatially shuffled versions to ensure the same degree of perturbation in adversarial and ``control'' examples. Spatial shuffling was performed by randomly permuting perturbation values for all pixels.

\subsection{Pre-training on ImageNet}
In this set of experiments, we measured the attack effectiveness when target and surrogate were both pre-trained on ImageNet, both randomly initialized, or had different initializations (pre-trained or random). We measured this for target-surrogate pairs with the same and different architectures separately.
For this purpose, we trained four versions of each architecture (two pre-trained and two randomly initialized) to cover all possible target-surrogate combinations in black-box settings, using the same partition of the development set, \emph{d1}.

\subsection{Development data disparity}
This experimental setup focused on the effect of disparity in the data used for the development of the target and surrogate models, as well as its interaction with architecture disparity. \revised{For the first part of this set of experiments}, we trained four randomly initialized versions of each architecture: a target model trained and validated on \emph{d1} and three surrogate models trained and validated on \emph{d1}, \emph{d2}, and \emph{d2/2}, respectively.
For every development dataset, the same split between training and validation images was used to train every model.

\revised{For the second part, we experimented with target models trained on small datasets attacked by surrogate models trained on larger, non-overlapping datasets. Since pre-training on ImageNet is often needed to reach good performance in models trained on small datasets, we have included it in this experiment. As target models, we trained pre-trained and randomly initialized versions of each architecture on \emph{d1/10}; as surrogate models, we trained pre-trained and randomly initialized versions of each architecture on \emph{d2}. For every development dataset, the split between training and validation images was the same as in previous experiments.}

\section{Results}
\subsection{Perturbation degree}
The results of our experiments with different attack methods (FGSM and PGD) at different perturbation degrees can be found in Table \ref{diffattacks}.
\revised{The results for individual models are included in the Supplementary Material.}
Increasing adversarial perturbation degree decreased the target model's performance in most cases.
The experiments with control noise (spatially shuffled noise) showed that in the ophthalmology and radiology datasets the decrease in the performance of the target could be partially attributed to image corruption. However, this effect was quite small, except for the FGSM attack in the ophthalmology dataset.
FGSM and PGD attacks performed similarly for the radiology and pathology dataset. For the ophthalmology dataset, the FGSM attack decreased the performance of the target model more than the PGD attack. We chose to use both attacks in our subsequent experiments and report average results.

\begin{table*}[t!]
\caption{Effects of perturbation degree on attack transferability. Average performance (AUC) over two model architectures is shown when using FGSM, PGD or control noise (spatially shuffled black-box adversarial perturbations) with varying perturbation degrees. The target and surrogate model were both trained with the same dataset, $d1$. \final{The lowest AUC value (highest attack transferability) in each application is shown in bold.}
}
\label{diffattacks}
\setlength{\tabcolsep}{4pt}
\centering
\begin{tabular}{c|c|c|c|c|c|c|c|c|c|c|c|c|c}
Data 
& Noise  
& \multicolumn{6}{c|}{FGSM} & \multicolumn{6}{c}{PGD} \\
\hline
\multicolumn{2}{r}{$\epsilon =$} & 0.01 & 0.02    & 0.03 & 0.04  & 0.05   & 0.06    &  0.01 & 0.02     & 0.03 & 0.04   & 0.05 & 0.06  \\

\hline
Ophthalmology     & 
None &
\multicolumn{12}{c}{0.86} \\
\hline
Ophthalmology       
& Adversarial & 0.56 & 0.44    & 0.37 &  \textbf{0.32}  & \textbf{0.32} & 0.33   & 0.72 & 0.56     & 0.44  & 0.37  & 0.35 & 0.34      \\
Ophthalmology       
& Control     & 0.85 &  0.85   & 0.84 & 0.79     & 0.76 & 0.73   & 0.86 & 0.85    & 0.85 & 0.84  & 0.84 & 0.84      \\
\hline
Radiology   & 
None & 
\multicolumn{12}{c}{0.75} \\
\hline
Radiology  
& Adversarial    
 &0.61 & 0.55  & 0.52 & 0.51  & 0.51 & 0.52  & 0.65 & 0.57  & 0.52 & 0.49  & 0.47 & \textbf{0.45} \\
Radiology  
& Control  & 0.75 & 0.75  & 0.75 & 0.74  & 0.73 & 0.72  & 0.75 & 0.75  & 0.75 & 0.75  & 0.74 & 0.74  \\
\hline
Pathology    & 
None & 
\multicolumn{12}{c}{0.87} \\
\hline
Pathology      
& Adversarial
 & 0.70 & 0.56  & 0.45 & 0.38  & 0.35 & \textbf{0.33}  & 0.73 & 0.56  & 0.47 & 0.41   & 0.38 &  0.36    \\
Pathology       
& Control     & 0.87 & 0.87   & 0.87 & 0.87  & 0.87 & 0.87  & 0.87 & 0.87   & 0.87 & 0.87  & 0.87 &   0.87  \\
\end{tabular}
\end{table*}

\begin{figure*}[t!]
\centering
\includegraphics[width=\textwidth]{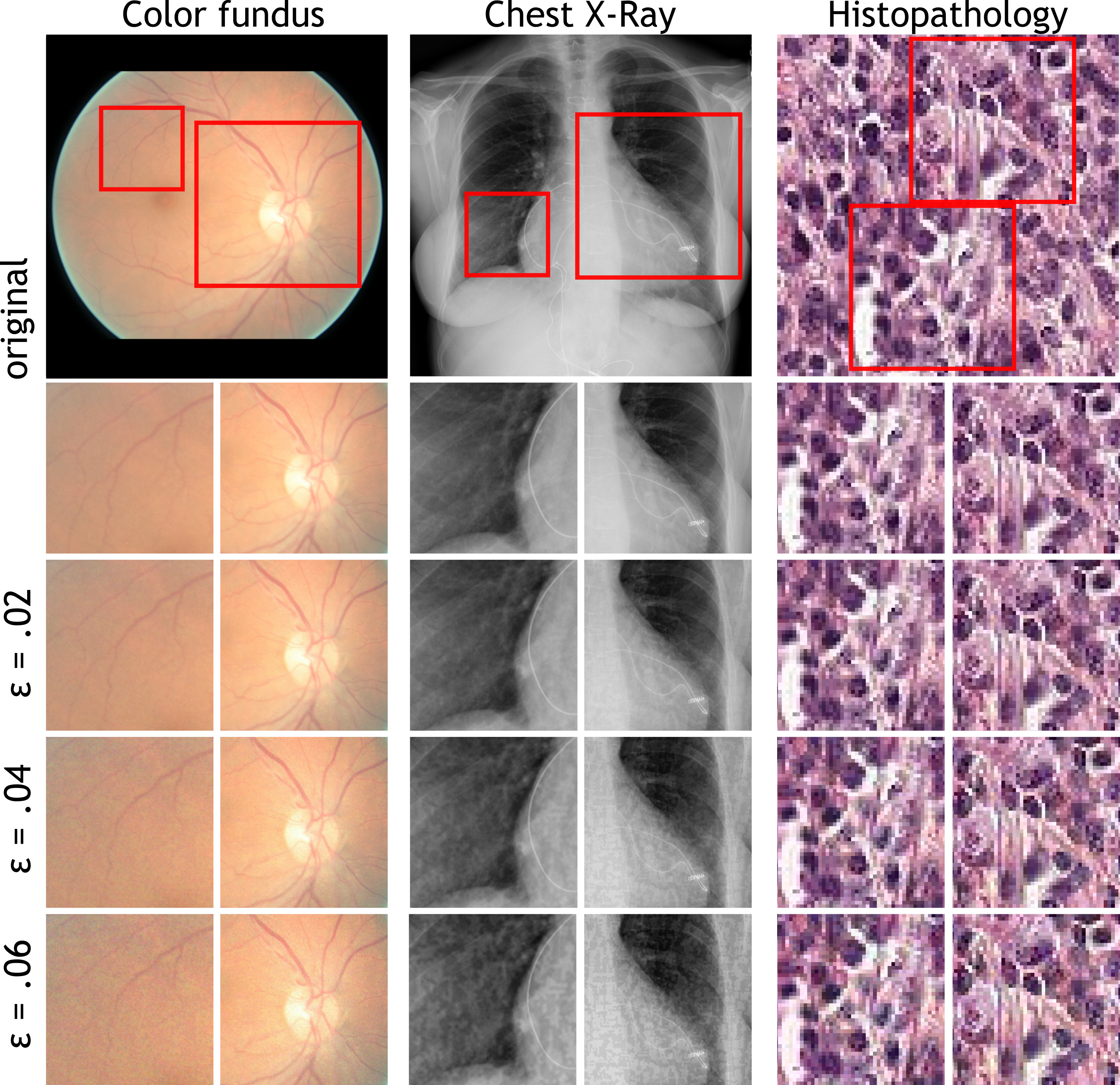}
\caption{
Original and adversarial images created with fast gradient sign method attacks using different perturbation degrees ($\epsilon$). The images in the top row are the original images. The red squares indicate the location of the patches that we show in the rest of the figure.
}
\label{fig_perceptibility}
\end{figure*}

\begin{figure*}[h!]
	\begin{center}   		
		\includegraphics[width=\linewidth]{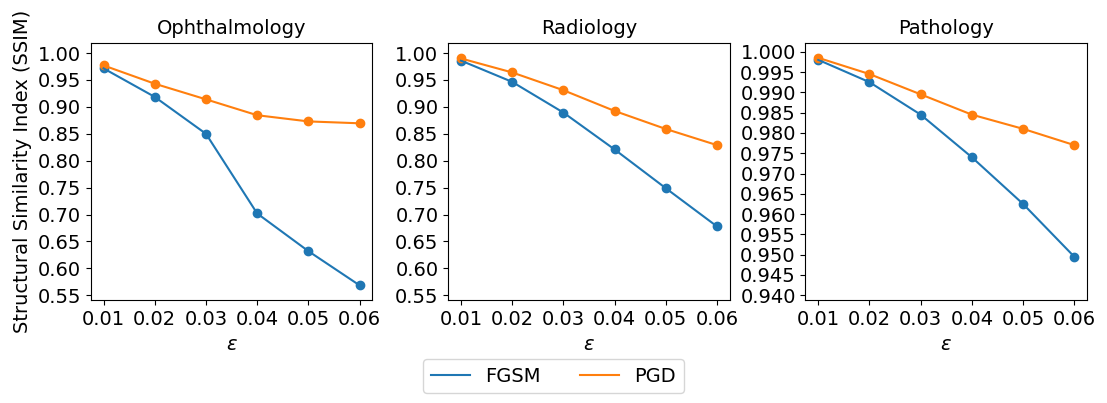}
	\end{center}
	\caption{\revised{Mean Structural Similarity Index Measure (SSIM) computed between original images in the test sets and adversarial examples generated using FGSM or PGD with varying perturbation degree $\epsilon$. Points represent SSIM values averaged over two model architectures (Inception-v3 and Densenet-121).}}
\label{figure:fig_ssim} 
\end{figure*}

Figure \ref{fig_perceptibility} shows original images and their adversarial counterparts computed using FGSM attacks at different perturbation degrees.
\revised{Figure \ref{figure:fig_ssim} shows mean SSIM values across all images for FGSM and PGD attacks.}
\revised{SSIM values for individual models are included in the Supplementary Material.}
As can be seen, applying the same amount of perturbation to different imaging modalities has a different effect on \revised{human visual perceptibility and the measured SSIM. Adversarial perturbations were the most noticeable in the radiology images, with $\epsilon = 0.02$ yielding an already visible, albeit quite subtle perturbation. For the ophthalmology and pathology images, at the same perturbation degree, perturbations were almost imperceptible and became noticeable with higher epsilon values. Perturbations computed by FGSM had lower SSIM than those computed by PGD in all three datasets. This is an expected result, since PGD optimizes perturbations according to both their impact on model predictions and their size.}

For our further experiments, we chose to report attacks using $\epsilon = 0.02$, as this was the highest perturbation degree that was still visually subtle for all applications \revised{and attack methods, and it had substantially better transferability than an epsilon of 0.01 in most of the studied applications}.

\subsection{Pre-training on ImageNet}
Table \ref{pretraining} summarizes our experiments on the effect of pre-training on adversarial attack transferability and its interaction with model architecture parity. 
\revised{The results for individual models and different attack methods can be found in the Supplementary Material.}
In the ophthalmology and radiology datasets, the attack transferability between pre-trained models was substantially higher than that between randomly initialized models.
In both datasets, the effect was consistent: for all eight combinations of attack method and target and surrogate pairs (including pairs having a different architecture), pre-trained targets had lower performance when attacked by pre-trained surrogates, compared to their randomly initialized counterparts. In the pathology dataset, however, the opposite effect was observed with similar consistency.
It is noteworthy that the effect of pre-training on transferability seemed to correlate to the performance increase resulting from pre-training: in the ophthalmology dataset, both the performance boost obtained by using pre-training and the transferability of adversarial examples between pre-trained networks were high; in the radiology dataset, the performance boost was smaller and the transferability was also smaller; in the pathology dataset, pre-training yielded no benefit and the effect on transferability was reversed.

\begin{figure*}[h!]
	\begin{center}   		
		\includegraphics[width=\linewidth]{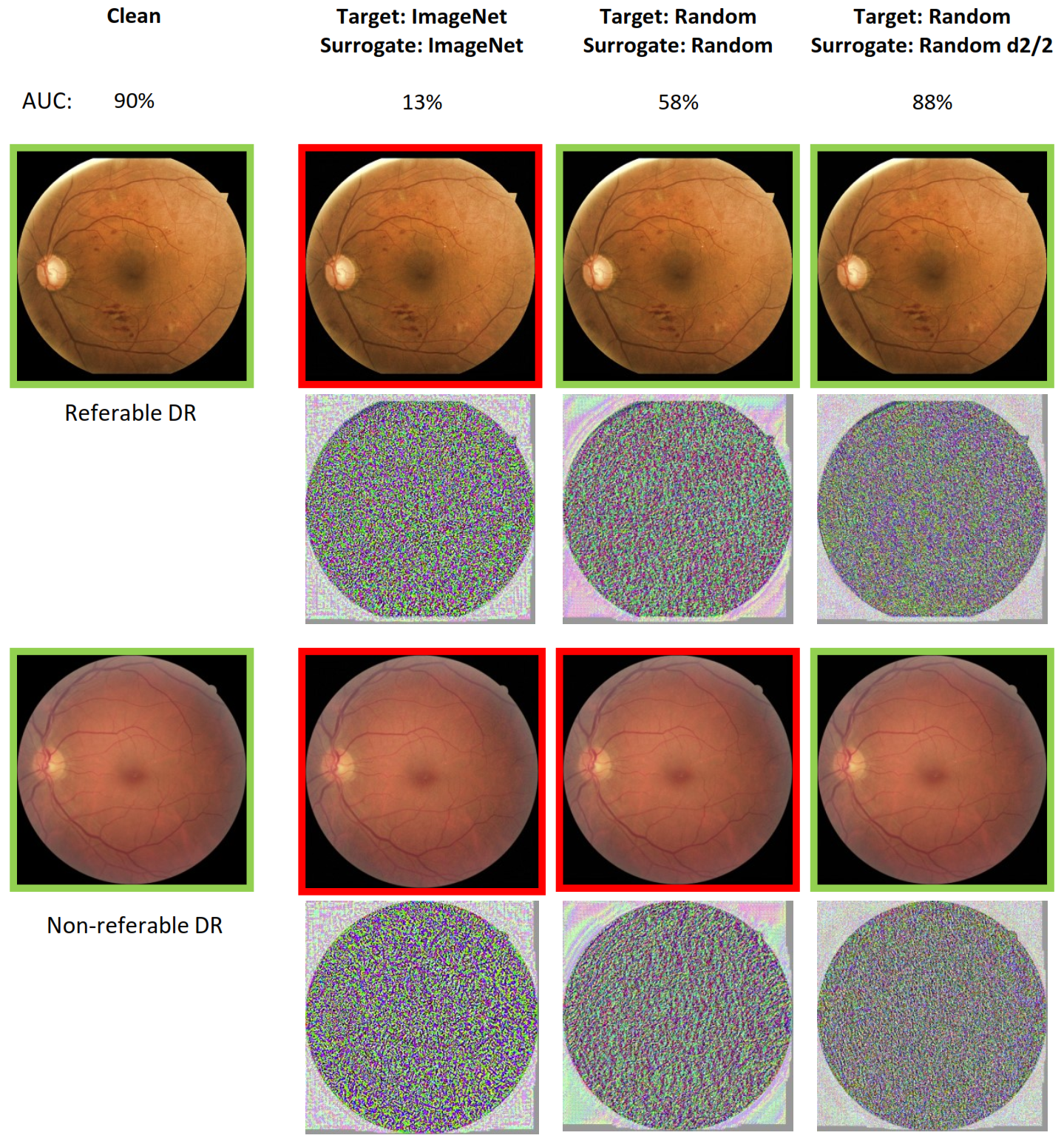}
	\end{center}
	\caption{Original images, adversarial images and corresponding adversarial noise created with FGSM ($\epsilon$=0.02) in different black-box settings: target and surrogate pre-trained on ImageNet; target and surrogate randomly initialized; target and surrogate randomly initialized plus surrogate developed using a different and reduced dataset (d2/2). The average area under the receiver operating characteristic curve (AUC) is indicated above of each configuration for the clean and the black-box settings. Green frame indicates correct classification of referable or non-referable diabetic retinopathy (DR); red frame, incorrect classification. \revised{The adversarial noise shown is equivalent to the difference between the original and the adversarial image.}}	
\label{figure:fig_experiments} 
\end{figure*}

Figure \ref{figure:fig_experiments} includes two examples from the ophthalmology dataset that illustrate attack transferability when both target and surrogate are pre-trained on ImageNet and when both are randomly initialized.

All the aforementioned effects held similarly for the scenarios where the target and surrogate model had the same or different architecture.

\begin{table*}[t!]
\caption{Effects of pre-training and the interaction between pre-training and model architecture parity on attack transferability. Average performance (AUC) over FGSM and PGD ($\epsilon$=0.02) and two model architectures (Inception-v3 and Densenet-121)
is shown. The target and surrogate model were both trained with the same dataset, $d1$. Average relative performance with respect to the no-attack setting is shown in brackets. \final{The lowest AUC value (highest attack transferability) in each application is shown in bold.}}
\label{pretraining}
\setlength{\tabcolsep}{4pt}
\centering
\begin{tabular}{c|c|c|c|c|c}
Architecture & Target   & Surrogate & Ophthalmology               & Radiology & Pathology \\
\hline
No attack            & Imagenet & -         & 0.94 (100\%)                        & 0.78 (100\%)     & 0.87 (100\%)     \\
No attack            & Random   & -         & 0.86 (100\%)                       & 0.75 (100\%)     & 0.87 (100\%)     \\
\hline
Same         & Imagenet & Imagenet  & \textbf{0.00 (0\%)} & \textbf{0.31 (40\%)}      & 0.61  (70\%)    \\
Same         & Random   & Random    & 0.44 (51\%)                        & 0.48 (64\%)      & \textbf{0.41 (47\%)}     \\
Same         & Random   & Imagenet  & 0.63 (74\%)                       & 0.63 (83\%)      & 0.60 (69\%)     \\
Same         & Imagenet & Random    & 0.80 (85\%)                       & 0.55 (71\%)      & 0.71 (82\%)     \\
\hline
Different    & Imagenet & Imagenet  & 0.24 (25\%) & 0.50 (65\%)     & 0.75  (86\%)    \\
Different    & Random   & Random    & 0.55 (64\%)                       & 0.64 (86\%)      & 0.71 (82\%)\\
Different    & Random   & Imagenet  & 0.71  (83\%)                      & 0.65 (86\%)      & 0.69  (80\%)    \\
Different    & Imagenet & Random    & 0.86 (92\%)                       & 0.59 (76\%)      & 0.75  (86\%)    \\
\end{tabular}
\end{table*}

\subsection{Development data disparity}
The effects of data disparity on adversarial attack transferability and its interaction with model architecture disparity can be seen in Table \ref{diffdata}.
\revised{The results for individual models and different attack methods are included in the Supplementary Material.}
For all datasets, networks were substantially less susceptible to attacks crafted using surrogates with the same architecture but trained on a different data subset (\emph{d2} or \emph{d2/2}).
This held for both target architectures and both attack methods.
Decreasing the surrogate training set size (from \emph{d2} to \emph{d2/2}) led to a further drop in the attack transferability for the ophthalmology and radiology datasets.

When the architecture of the surrogate was different, however, additional data disparity between the target and surrogate substantially decreased the attack performance only for the ophthalmology dataset.
Disparity in the model architecture had greater effect on attack performance than disparity in data for the radiology and pathology datasets; for the ophthalmology dataset, data and model architecture disparity had similar effects.

\revised{The transferability of attacks on models trained on small datasets in a data disparity scenario is reported in Table \ref{diffdata2}. \revised{The Supplementary Material contains the results for individual models and different attack methods.} For the ophthalmology and radiology datasets, the pre-trained models clearly outperformed their randomly initialized counterparts on clean images. For the pathology dataset, pre-trained models performed slightly worse than randomly initialized ones. These results are similar to ones we observed for models trained on larger sets (see Table \ref{pretraining}). On adversarial images, pre-trained models performed worse than their randomly initialized counterparts in all three datasets, both in absolute terms and relative to their performance on clean images. These results were mostly similar to the results for networks trained on larger data (Table \ref{pretraining}). For the ophthalmology and radiology datasets, adversarial attack transferability between pre-trained models was higher than that between randomly initialized models, and this effect was stronger in the ophthalmology dataset. There was an interesting difference, however: for the pathology dataset, pre-training increased transferability, whereas in our experiments with networks trained on larger data it was the other way around. Attacks on randomly initialized models trained on small datasets hardly have any effect (Table \ref{diffdata2}), while attacks on randomly initialized models trained on larger sets can lead to performance decreases of up to 35\% (Table \ref{diffdata}).}

\begin{table*}[t!]
\caption{Effects of data and model architecture parity on attack transferability. Average performance (AUC) over FGSM and PGD ($\epsilon$=0.02) and two model architectures is shown, with surrogate models trained on different sets while the target model is trained on $d1$. Average relative performance with respect to the no-attack setting is shown in brackets. \final{The lowest AUC value (highest attack transferability) in each application is shown in bold.}}
\label{diffdata}
\setlength{\tabcolsep}{4pt}
\centering
\begin{tabular}{c|c|c|c|c}
Architecture & Training set & Ophthalmology & Radiology & Pathology \\ \hline
No attack               & -               & 0.86 (100\%)         & 0.75 (100\%) & 0.87 (100\%)     \\ \hline
Same            & d1              & \textbf{0.44 (51\%)} & \textbf{0.48 (64\%)} & \textbf{0.41 (47\%)}     \\
Same            & d2              & 0.56 (65\%) & 0.56 (75\%) & 0.67 (77\%)     \\
Same            & d2/2              & 0.75 (88\%) &0.59 (79\%) & 0.65 (75\%)     \\ \hline
Different       & d1              & 0.55 (64\%) &0.64 (86\%) & 0.71 (82\%)     \\
Different       & d2              & 0.66 (77\%) &0.65 (87\%) & 0.74 (85\%)     \\
Different       & d2/2              & 0.80 (93\%) &0.69 (91\%) & 0.71 (81\%)    
\end{tabular}
\end{table*}

\revised{\begin{table*}[t!]
\caption{\revised{Transferability of attacks on models trained on small datasets in a data disparity scenario. Average performance (AUC) over FGSM and PGD ($\epsilon$=0.02) and two model architectures is shown, with surrogate models trained on $d2$ and target models trained on $d1/10$. Average relative performance with respect to the no-attack setting is shown in brackets.} \final{The lowest AUC value (highest attack transferability) in each application is shown in bold.}}
\label{diffdata2}
\setlength{\tabcolsep}{4pt}
\centering
{\begin{tabular}{c|c|c|c|c|c}
Architecture & Target & Surrogate & Ophthalmology & Radiology & Pathology \\ \hline
No attack               & Imagenet & -               & 0.88 (100\%)         &  0.69 (100\%) &  0.79 (100\%)     \\
No attack               & Random & -               & 0.61 (100\%)         &  0.64 (100\%) &  0.81 (100\%)     \\ \hline

Same            & Imagenet & Imagenet              & \textbf{0.13 (15\%)} & \textbf{0.58 (84\%)} & \textbf{0.73 (93\%)}     \\
Same            & Random & Random              & 0.60 (99\%) & 0.63 (99\%) &  0.78 (96\%)     \\ \hline
Different            & Imagenet & Imagenet             & 0.45 (51\%) & 0.62 (90\%) &  0.75 (96\%)     \\ 
Different       & Random & Random             & 0.60 (100\%) & 0.63 (99\%) &  0.79 (98\%)     \\
\end{tabular}}
\end{table*}}

\section{Discussion}
In this study, we have demonstrated that ImageNet pre-training may substantially affect transferability of adversarial examples, even between networks with different architecture. This effect varied substantially across the applications and appeared to be related to the gain in performance resulting from pre-training. We have also shown that differences in development data between target and surrogate models reduce transferability substantially, even when development sets are equally sized and sampled from the same distribution. This effect was in some cases comparable to that of architecture disparity. All experiments were performed using a perturbation degree tuned to be visually subtle and perform optimally in the black-box attack setting.

In this section, we discuss the importance of perturbation degree tuning and the influence of pre-training and data disparity on transferability of adversarial attacks. Based on the results of our study, we make recommendations for developers of MedIA systems, as well as for future evaluation of adversarial robustness of these systems.

\subsection{Perturbation degree}

Our experiments confirmed that perturbation degree is an important attack parameter to take into account to obtain more accurate estimates of adversarial robustness of DL systems. Using a lower-than-optimal perturbation degree may lead to an underperforming attack and hence an overestimated robustness; using a higher-than optimal perturbation degree may make the adversarial perturbation visually perceptible. 
We observed differences in visual perceptibility of adversarial perturbations in different imaging modalities \revised{as estimated by both our visual perception and SSIM}. This could occur because of differences in color, homogeneity, contrast, and resolution between the imaging modalities.
Since these characteristics may affect visual perceptibility of adversarial attacks, it is important to optimize the perturbation degree to the image type and task, as well as for the considered attack scenario (e.g., whether adversarial examples are likely to be inspected by a human). \revised{Quantitative measures, such as SSIM, could also be used to ensure minimal visual perceptibility of adversarial perturbations. However, since there is not an accepted threshold value to determine whether a perturbation is imperceptible for SSIM or other quantitative perceptibility measures, an optimal threshold value would still need to be found and ensured to agree with human visual perception. Moreover, to our knowledge, no quantitative metric can perfectly capture human visual perceptibility \citep{chandler2013seven}. Therefore, we think the best way to assess visual perceptibility of different degrees of adversarial perturbations would be a blinded observer study involving medical experts. Such a study is beyond the scope of this paper.}

\cite{ma2021understanding} experimented with different perturbation degrees in the white-box attack setting and concluded that MedIA systems are ``easier to attack'' than systems trained on natural images, based on their observation that for MedIA systems far smaller perturbations were needed to reach near-maximal attack performance. In our study, we considered the more realistic black-box setting, in which perturbations became visually perceptible before yielding high attack performance. This suggests that, firstly, in black-box settings, MedIA systems may not be very easy to attack.
Secondly, it suggests it is harder to compare the difficulty of attacking systems in different applications: for example, in applications where a given perturbation degree yields better attack effectiveness, the perturbations may also be more perceptible. 

\subsection{Pre-training on ImageNet}

In the ophthalmology and radiology applications, we observed that transferability between pre-trained models, including the ones with different architectures, was substantially larger than that between randomly initialized models: 20-50\% difference in AUC was observed.
These results motivate caution in generalizing performance of black-box adversarial attacks from pre-trained networks to randomly initialized ones and vice versa. For example, an attack that was only shown effective on pre-trained targets and surrogates  may be substantially less effective when applied to randomly initialized targets and surrogates in the same application or to networks in applications that do not benefit from pre-training.

We believe increased transferability between pre-trained models may be explained by increased closeness of their decision boundaries. \cite{tramer2017space} showed empirically that decision boundaries of DL models are on average closer to each other than to data points, which implies that adversarial perturbations causing data points to cross one model's decision boundary would likely cause them to cross another model's decision boundary as well. There are several possible mechanisms through which pre-training may increase closeness of decision boundaries of models. Firstly, pre-trained networks with the same architecture start with the same weight initialization (whereas randomly initialized networks in our experiments started with different initializations), which may increase the similarity of the features they learn. The fact that pre-training speeds up convergence may amplify this. Secondly, pre-trained networks may be more similar because they retain some features from ImageNet pre-training. 
As, in our experiments, pre-training also increased transferability between models with different architectures, same weight initialization is likely not the only cause of increased similarity between pre-trained networks. The correlation between the strength of the performance boost from pre-training and the increase in transferability also supports the second mechanism: the higher the performance gain from pre-training, the more the network retains from its ImageNet pre-training.

\revised{Our observations put into an interesting perspective the ones made in the study by \cite{hendrycks2019using} — the only study on the effects of pre-training on adversarial robustness we are aware of. \cite{hendrycks2019using} found that adversarial pre-training on ImageNet can increase adversarial robustness for networks adversarially fine-tuned on the target data in the white-box attack setting. We found that regular ImageNet pre-training can decrease adversarial robustness in the black-box setting. Whether adversarial pre-training could instead improve robustness in the black-box setting remains an open question. On the one hand, adversarial training is substantially less successful in preventing attacks in the black-box than in the white-box setting \citep{tramer2017ensemble} and these results could be expected to extend to adversarial pre-training. Furthermore, if any kind of pre-training increases vulnerability to black-box attacks by similarly pre-trained networks, for example, by increasing similarity between the decision boundaries of the target and the surrogate, adversarial pre-training could be less beneficial or even detrimental to black-box robustness. On the other hand, even if adversarial pre-training facilitated transferability to some degree, it could still be overall beneficial due to the fact that the network would be trained to be adversarially robust on a larger and more variable set of images. Future research could focus on answering these questions.}

\subsection{Development data disparity}

Data parity is assumed, to our knowledge, by all studies on black-box adversarial attack robustness of MedIA systems \citep{finlayson2018adversarial, taghanaki2018vulnerability,paschali2018generalizability}. Our results, however, indicate that black-box attacks may be less effective when using a surrogate trained on a different dataset, even if it is a large dataset of the same size as the development data of the target and it is sampled from the same distribution. A 30-40\% increase in AUC of the attacked models in the ophthalmology and pathology datasets was observed when the surrogate was trained on a disjoint subset. This data disparity effect may be as strong or even stronger than the effect of architecture disparity, as observed in the ophthalmology dataset.

\revised{The data disparity effect is even stronger when the target model is trained on a small dataset, in which case attacks are generally quite ineffective, especially when performed against randomly initialized target models. However, pre-trained models trained on small datasets can still be vulnerable to adversarial attacks. This vulnerability increases for applications where ImageNet pre-training provides a significant boost in clean performance, similar to what was observed in the experiments with models trained on larger datasets.}

These results suggest that MedIA systems that use private development data \revised{are less susceptible} to adversarial attacks \revised{than systems that use public development data (assuming attacks performed by an external party who cannot access the private data and assuming other properties of the systems are equal)}. Simulating data disparity between the target and surrogate model yields a more realistic estimate of adversarial robustness for such systems. Studies on adversarial robustness would therefore benefit from including different attack scenarios assuming data parity and disparity\revised{, including differences in development data sizes of target and surrogate networks,} in their evaluation.

\subsection{\final{Adversarial robustness: Inception-v3 vs Densenet-121}}
\final{Considering the results included in the Supplementary Material for each implemented model architecture, Inception-v3 and Densenet-121, we observed that, in the ophthalmology application, target models based on Inception-v3 tended to be more vulnerable when attacked by surrogate models with the same architecture, whereas target models based on Densenet-121 were slightly more vulnerable when attacked by surrogate models based on Inception-v3 (compared to Inception-v3 attacked by Densenet-121 models). In the radiology application, target models based on Inception-v3 were observed to be on average more vulnerable than those based on Densenet-121, although no substantial differences were observed for ImageNet pre-trained versions of the models. In the pathology application, target models based on Densenet-121 were found to be slightly more vulnerable in most scenarios. Furthermore, when there was development data disparity between target and surrogate models, only small differences in robustness between architectures were observed in all applications. 

\cite{su2018robustness} studied transferability of adversarial attacks between popular architectures trained on ImageNet. Densenet-121 was observed to be more robust, often substantially, to FGSM and PGD attacks by Inception-v3 than the other way around. Simultaneously, there was almost perfect transfer between different variants of Densenet: Densenet-121, Densenet-161, and Densenet-169 (although transferability between the same version of architectures were not reported). Our results showed different trends when comparing Densenet-121 and Inception-v3 in different applications, for different weight initializations (ImageNet pre-training or random initialization), and for different target-surrogate development data configurations. For example, we observed perfect or high transferability between Densenet-121 models only for ophthalmology and radiology applications and only for the ImageNet-pretrained versions. It is thus difficult to conclude whether either of these architectures is innately more robust to black-box attacks than the other.
}

\subsection{Recommendations for developers of MedIA systems}

We recommend developers of all MedIA systems to be deployed in clinical practice to consider the environment their system will be used in and assess whether the following holds:
\begin{enumerate}
\item Users of these systems may have a motivation (financial or otherwise) to manipulate their output.
\item Users may have the capacity to manipulate their input without being detected.
\end{enumerate}
For MedIA systems satisfying these criteria, especially those systems that significantly affect clinical or financial decision-making, we recommend taking proactive measures to mitigate the risk of successful adversarial attacks.

Many different methods have been proposed to defend DL systems from adversarial attacks \citep{yuan2019adversarial,akhtar2018threat,biggio2018wild}.
\revised{Although all defense methods proposed to date are only partially effective \citep{yuan2019adversarial}, applying the most successful methods is likely to increase the difficulty of manipulating DL systems. We thus recommend developers of security-critical MedIA systems to consider employing some of these strategies. In addition to strategies purposefully designed to defend against adversarial attacks, quantifying uncertainty and using techniques for interpreting predictions may aid in detecting adversarial attacks \citep{li2017dropout,tao2018attacks}. It was shown that adversarial perturbations can increase the model's uncertainty \citep{li2017dropout} and cause discrepancies in interpretations of the model's predictions \citep{tao2018attacks}. However, detection of adversarial examples based on uncertainty and interpretability also provides only partial protection against adversarial attacks \citep{carlini2019ami, smith2018understanding}, and can be easily circumvented when taken into  account in the attack method \citep{zhang2020interpretable}.}

\revised{Given that adversarial defense methods are not fully reliable, and given that increased transferability between similar models was observed in this and other studies (for example, \cite{su2018robustness}), we also recommend taking measures to increase the difficulty of training a surrogate model similar to the target.}
As one such measure, we recommend restricting the amount of information on the design of the system available to the public. This includes information on the methodology components of the system, such as network architecture and weight initialization. We also recommend avoiding disclosing extensive details on the system’s data: for example, names and identifying details of used public data, detailed information on distribution of subjects, scanning modalities, and protocols. However, we do not recommend keeping secret the methods, procedures, and description of datasets used to evaluate the system, since this would make it harder to ensure the system is safe and has the desired performance level.

To further increase the difficulty of emulating the target model for an attacker, we recommend considering re-designing MedIA systems to reduce the use of standard components, such as popular network architectures, and components that facilitate transferability, such as pre-training, as well as reducing the reliance of these systems on publicly available development data. For example, standard architectures could be replaced by customized architectures and pre-training may be substituted by random initialization. However, we recommend this strategy only as a complement to more explicit defense strategies and only if it does not lead to a significant decrease in the system performance or substantially slow down its development.

\revised{
We acknowledge that our recommendation to avoid using standard components, such as pre-training on ImageNet and publicly available development datasets might hamper performance.  However, for  MedIA  systems  planned  to  be  deployed  in  clinical practice, robustness needs to be considered in addition to performance. The  trade-off between  performance  and  robustness has already been discussed by others \citep{zhang2019theoretically, tsipras2018robustness, paschali2018generalizability}. The decision on how much performance to sacrifice for robustness will differ per case depending on the likelihood of adversarial attacks against the given system and their potential consequences.
}

\revised{We believe a combination of several defense strategies would provide the most comprehensive security. Thus, we recommend combining multiple methods for detecting, neutralizing, or robustifying against adversarial perturbation, with measures that increase the difficulty of modeling the target system for a potential attacker.}

We would like to emphasize that all recommendations above only apply to systems that are planned to be deployed in practice. \revised{However, we believe they are also relevant to researchers developing  models at earlier stages, or performing research not specifically focused on adversarial attacks. We consider it important that MedIA researchers are aware of the effect that commonly used design components, such as pre-training on ImageNet or public datasets, have on attack transferability and the existing trade-off between performance and robustness of DL systems. This way, researchers will acknowledge the role of adversarial vulnerabilities in model development, with the capability of “shifting” what is currently standard in MedIA towards components that acknowledge these vulnerabilities as well.}

\subsection{Recommendations for evaluating adversarial robustness of MedIA systems}
\cite{carlini2019evaluating} presented a detailed discussion on best evaluation practices to conduct reproducible, falsifiable studies on adversarial robustness of DL systems. They place an emphasis on estimating the upper bound of adversarial robustness: that is, adversarial robustness measured against attacks of the maximally knowledgeable and capable attacker. Below is a condensed list of their general recommendations:

\begin{itemize}
\item State a precise threat model that the target system is supposed to be robust under.
\item Perform adaptive attacks to estimate the upper bound of robustness: test attacks that have full access to the defense mechanisms the target system might use and adapt attacks to the target system so as to maximize their effectiveness. This includes carefully investigating the attack parameters to ensure optimal attack performance.
\item Perform various sanity checks on the success rates of the attacks to ensure they are correctly implemented and their methodology is valid (for example, white-box iterative attacks should perform better than one-step attacks; attacks adapted to the studied system should perform at least as good as any other).
\item Test diverse attacks (e.g. one-shot attacks and iterative attacks).
\item Describe the attacks studied fully, including parameters.
\item Compare against prior work and explain important differences.
\end{itemize}

Current studies on adversarial attacks on MedIA systems do not follow all of these practices. To the best of our knowledge, no published studies investigating robustness of MedIA systems formulate an explicit threat model, and thus do not clearly define the considered attack scenarios; many do not tune the parameters of their attacks, including perturbation degree \citep{paschali2018generalizability,taghanaki2018vulnerability,finlayson2018adversarial}; and some do not report all attack parameters \citep{paschali2018generalizability,taghanaki2018vulnerability}.

Inspired by the results in our study, we have developed several additional recommendations and suggestions for evaluating adversarial robustness of DL systems. Note that while recommendations of \cite{carlini2019evaluating} (particularly the recommendation on performing adaptive attacks) have as their aim estimating the upper bound of adversarial robustness, our recommendations have a different scope. We aim at investigating factors that may affect adversarial vulnerability of real-world DL systems, which are unlikely to be completely known by the attacker, as well as at obtaining realistic estimates of robustness of such systems.

\begin{itemize}
\item For image analysis (including MedIA) applications, we recommend tuning the perturbation degree ($\epsilon$ or another parameter controlling it) to the target image type or modality, so that the attack yields maximal performance while the perturbations still satisfy a chosen criterion for bounding perturbation degree, such as  visual perceptibility. Such criterion should be explicitly defined and measured. For example, if the criterion is visual perceptibility, we suggest the studies to describe how perceptibility was judged and to provide fully-sized or zoomed-in versions of images that the reader can also examine.
\item We encourage researchers to consider design components shared by both target and surrogate that may increase the similarity between them and study the effect of changing such settings on attack transferability. For example, pre-training on ImageNet, development data parity, and architecture parity could be considered as similarity-promoting components as in our study. Other similarity-promoting settings could focus on regularization techniques, which encourage networks to have specific properties (e.g. weight decay, deformation consistency regularization), loss function, pre-processing, data augmentation protocol, \revised{or popular network architectures other than the ones we used and their properties (such as ResNets and skip connections, found to increase adversarial vulnerability  \citep{wu2020skip})}.
\end{itemize}

The recommendations developed by \cite{carlini2019evaluating} and by us are aimed at public scientific studies on adversarial robustness. However, we can envision a different setting for evaluating robustness of DL systems where most of this advice may also be useful: a private evaluation setting in which the robustness of a closed-source DL system is evaluated by the developing company or a different organization. In this setting, it may be of interest to estimate robustness under the most likely attack scenarios, which may exclude scenarios where the attacker has complete or very comprehensive knowledge of the target system. Therefore, recommendations aimed at obtaining realistic robustness estimates, as opposed to the upper bound estimates, may be the most applicable. Recommendations we would not advise to apply in this setting are those concerning public disclosure of robustness evaluation procedure, including tested attacks and their parameters.

\section{Conclusion}
In this paper, we studied the influence of two previously unexplored factors on the transferability of black-box adversarial attacks in three different MedIA applications. We observed that pre-training on ImageNet may dramatically increase the transferability of adversarial examples in MedIA systems; the larger the performance gain achieved by pre-training, the larger the transfer and thus the more vulnerable the pre-trained system is to attacks by pre-trained surrogate models.
We also showed that disparity in development data and model architecture between target and surrogate models can substantially decrease the success of attacks. 
We believe these factors should be considered in the design of security-critical MedIA systems, especially those planned to be deployed in clinical practice. In order to reduce the transferability of potential attacks, \revised{in addition to using techniques developed for defending DL models against adversarial attacks,} we recommend restricting the disclosure of information on design specifications, as well as considering reducing the use of standard components (such as pre-training on ImageNet and popular network architectures) and publicly available datasets for development. Finally, we believe future studies on adversarial robustness of MedIA systems may benefit from simulating various attack scenarios and target-surrogate disparities. This may facilitate a better understanding of attack transferability and the factors that determine it, as well as more realistic robustness estimates for MedIA systems.

\section*{Acknowledgments}
This work was supported by the Deep Learning for Medical Image Analysis (DLMedIA) research program by The Dutch Research Council (project number P15-26), Intel Corporation (GB, IK, LH), and Philips Research (SCW, JPWP, MV). 

\bibliographystyle{model2-names.bst}\biboptions{authoryear}
\bibliography{bib}



\end{document}